# Bernoulli process with 282 ky periodicity is detected in the R-N reversals of the earth's magnetic field

Jozsef Garai
Department of Earth Sciences
Florida International University
University Park, PC-344
Miami, FL 33199
Ph: 305-348-3445
Fax: 305-348-3070
E-mail: jozsef.garai@fiu.edu

**ABSTRACT**

Investigating the polarity time scales for the last 118 My a Bernoulli process with $p=\frac{1}{2}$ overriding with a Gaussian noise have been detected for the R-N magnetic reversals. The Bernoulli trials are separated by 282 ky. The frequency of the Bernoulli process correlates to planetary cycles and consistent with an impact triggered global cooling mechanism. The detected pattern allows calculating the probability of an upcoming R-N reversal between any time intervals.

## I. INTRODUCTION

It has been proposed that the two kinds of magnetic reversals, normal to reverse (N-R), and reverse to normal (R-N) might be triggered by different mechanisms[1]. Global cooling could result an R-N reversal while the tidal affect and/or global warming could lead to an N-R reversal. If the two different kinds of magnetic reversals are triggered by different mechanisms then by looking for a pattern in the sequences of the reversals would more accurately be detected by focusing analysis on those reversals, which are triggered by a common mechanism. In the present study the periodicity of the R-N reversals will be investigated because these reversals are triggered by the same single mechanism.

Previous studies assumed that the mechanism responsible for the triggering of the magnetic reversal is the same in both cases. These studies were looking for sequences in the time series of the earth's magnetic field reversals by investigating the elapsed time between two consecutive reversals (see, e.g. Refs 2-3). The time sequence elapsed between the same polarity reversals has never been investigated.

## II. DATA ANALYZIS

Using the polarity time scales for the last 118 My[4] the time elapsed between two consecutive R-N reversals $\Delta t_{(R-N)}$ has been determined first.

$$\Delta t_{(R-N)} = t_{(R-N)_{i+1}} - t_{(R-N)_i}$$

where $t_{(R-N)_i}$ is the time before present of the $i^{th}$ R-N magnetic reversal. The number of data for longer time periods ($\Delta t_{(R-N)} > 1.6$ My.) is very few. These long periods were omitted because the available data is insufficient for statistical analysis. The frequency distribution of $\Delta t_{R-N}$ reveals five relatively distinct groups (Figure 1). If one assumes a Gaussian distribution for each of the groups, then these distributions overlap and the analyses are significantly more complicated. However, since the means of the various groups lay several standard deviations away from each other, a more tractable way to model the groups is to determine a point between them such that the probability of a point being miscategorized is minimized. To make this precise, consider $G_1$ and $G_2$ data set for two consecutive groups which have (mean, standard deviation) $(\mu_1, \sigma_1)$ and $(\mu_2, \sigma_2)$ respectively, and $\mu_1 < \mu_2$. Then we want to find $\mu_1 < s_{1-2} < \mu_2$ such that $p_1(X \geq s_{1-2}) + p_2(X < s_{1-2})$ is minimized, where $X$ is a variable equal to $\Delta t_{R-N}$, $p_1$ is a probability calculated from a Gaussian density function with parameters of $\mu_1, \sigma_1$, and $p_2$ is a probability calculated from a Gaussian density function with parameters of $\mu_2, \sigma_2$. At first the groups were separated by visual inspection and parameters $(\mu_i, \sigma_i)$ were calculated for each group ($G_i$). Using the calculated group parameters the minimum points $s_{i-(i+1)}$ between each consecutive group were determined. If this minimum point coincided with the initial assumption then the group separation was accepted. If the minimum point did not coincide with the initial separation then the above procedure was repeated using the new minimum point for the separation. Calculations were repeated as long as all the calculated $s_{i-(i+1)}$ coincided with the initial separation point. The calculated parameters of the separated groups are given in Figure 1. The means of the overlapping Gaussian distributions are $3.86\sigma$ apart, therefore, the overlapping data on each side of the distribution is less than 1.5 percent. This small percentage of overlapping does not have significant affect on the statistical parameters. Additionally, the cut off parts for groups 2, 3, and 4 are close to symmetrical further reducing the overlapping affect.

Analyzing the calculated statistical parameters, the mean values and standard deviations of these separated groups it was found that the mean values are integer multiples of the smallest mean, revealing a periodic pattern. If we let the normalize mean of a group denote the mean of that group divided by its group number (e.g. group 5's normalized mean is 0.28376 My), then we find that the normalized means range from 0.280 to 0.287 My. The weighted average of the means is 0.28214 My, the corresponding standard deviation is 0.073 My

The frequency distribution of the reversal time shows that the R-N reversal is more likely to have a shorter term than a longer one. Calculating the relative frequency that a reversal falls into a group it was find that the frequency for the first group with mean of 0.28 My is 0.5. The frequency of the second group (mean 0.56 My) is 0.21, while the frequency for the third group (mean 0.85 My) is 0.12 etc. (Figure 1). This is just a Bernoulli process with $p=\frac{1}{2}$ overriding with a Gaussian noise. This means that for every ~0.28 My, elapsed since the previous R-N reversal, there is approximately 50% chance that the earth's magnetic field would make a shift from a reversed to a normal orientation (Figure 2).

**III. CORRELATION WITH ASTRONOMICAL CYCLES**

The identified Bernoulli process has been stable at least for the length of the analyzed data set, which is about 100 My. With this time length only astronomical cycles are known to remain stable. Astronomical events; therefore, should play major role in the triggering of R-N reversals. This conclusion is consistent with the identified the presence of the planetary cycles in the strength and the inclination of the earth magnetic field for the past 2.25 My[5].

Geological observations indicates that the last two R-N reversals coincided with global cooling and major impacts.[6;7] The global climate oscillation is generated by the Milankovitch cycles[8]. The bit frequency of these cycles is consistent with the frequency of the Bernoulli trials. Impacts could be triggered by planetary perturbations. The beat frequency of the giant planets (93,418.3 y.)[9] is a harmonic of the Bernoulli trials. The $p=\frac{1}{2}$ probability of the Bernoulli process could be explained as positive and negative interferences between the Milankovitch and planetary cycles.

## IV. PROBABILITY OF AN R-N REVERSAL

The detected pattern and the statistical parameters allow one to predict the probability of the next R-N reversal between any time intervals. The probability of a Bernoulli trial is.

$$P(n) = p^n (1-p)^{1-n} \quad \text{for} \quad n = 0, 1 \tag{3.1}$$

where $p = \dfrac{1}{2}$. If zero of n means no R-N reversal, while one is equivalent with an R-N reversal then the probability of the $i^{th}$ R-N magnetic reversal is:

$$P(R-N)_i = p^n (1-p)^{1-n} = 0.5 \tag{3.2}$$

The uncertainty of the time of the Bernoulli trial is very high therefore the discrete Bernoulli trial should be replaced with a Gaussian distribution.

$$P(R-N)_i = p^n (1-p)^{1-n} \int_{-\infty}^{+\infty} P(t)_i \, dt = 0.5 \tag{3.3}$$

where $P(t)_i$ is the Gaussian density function for the $i^{th}$ trial.

$$P(t)_i = \dfrac{1}{\sigma\sqrt{2\pi}} e^{-\dfrac{(t+780-it_0)^2}{2\sigma^2}} \tag{3.4}$$

where t is the time from present in ky, $t_0$ is the mean, and $\sigma$ is the variance. The most recent R-N magnetic reversal, Matuyama-Brunhes, has been used as reference; therefore, 780 ky has been added to the current time. In order to calculate the probability of the occurrence of an R-N reversal between time $t_1$ and $t_2$ [$P(R-N)_{t_1-t_2}$] first the sequence number of the Bernoulli trial (i) should be identified. The sequence numbers can be determined from equation 3.5 by keeping only the integer parts of the calculated values.

$$\dfrac{t_1 + 780}{t_0} + 0.5 \quad \text{and} \quad \dfrac{t_2 + 780}{t_0} + 0.5 \tag{3.5}$$

If the beginning and the end of the investigated period falls into the same Bernoulli trial ($i_1 = i_2$) then the probability of the event can be calculated by integrating equation 3.3 between the time $t_1$ and $t_2$.

$$P(R-N)_{t_1-t_2} = p^n (1-p)^{1-n} \int_{t_1}^{t_2} \dfrac{1}{\sigma\sqrt{2\pi}} e^{-\dfrac{(t+780-it_0)^2}{2\sigma^2}} dt \tag{3.6}$$

The parameters, mean 282 ky, and variance 73 ky, determined by this investigation should be used for the calculation. If the time period covers two or more consecutive Bernoulli trials ($i_1 \neq i_2$) then the probability of the R-N reversal can be calculated as:

$$P(R-N)_{t_1-t_2} = \{1-[p^n(1-p)^{1-n}]^{(z+1)}\}\left[0.5 - \frac{\int_{-\infty}^{t_1}\frac{1}{\sigma\sqrt{2\pi}}e^{\frac{-(t+780-i_1 t_0)^2}{2\sigma^2}}dt + \int_{t_2}^{+\infty}\frac{1}{\sigma\sqrt{2\pi}}e^{\frac{-(t+780-i_2 t_0)^2}{2\sigma^2}}dt}{z+1}\right] \quad (3.7)$$

where $z = i_2 - i_1$.

**IV. CONCLUSIONS**

In this study the time sequence elapsed between two consecutive R-N reversals has been analyzed for the last 118 My. The pattern consistent with a Bernoulli process with $p=\frac{1}{2}$ overriding with a Gaussian noise. The time sequence between the Bernoulli trials is 282 ky. This frequency correlates to planetary cycles and consistent with an impact triggered global cooling reversal mechanism. The detected Bernoulli process allows one to calculate the probability of an upcoming R-N reversal between any given time interval.

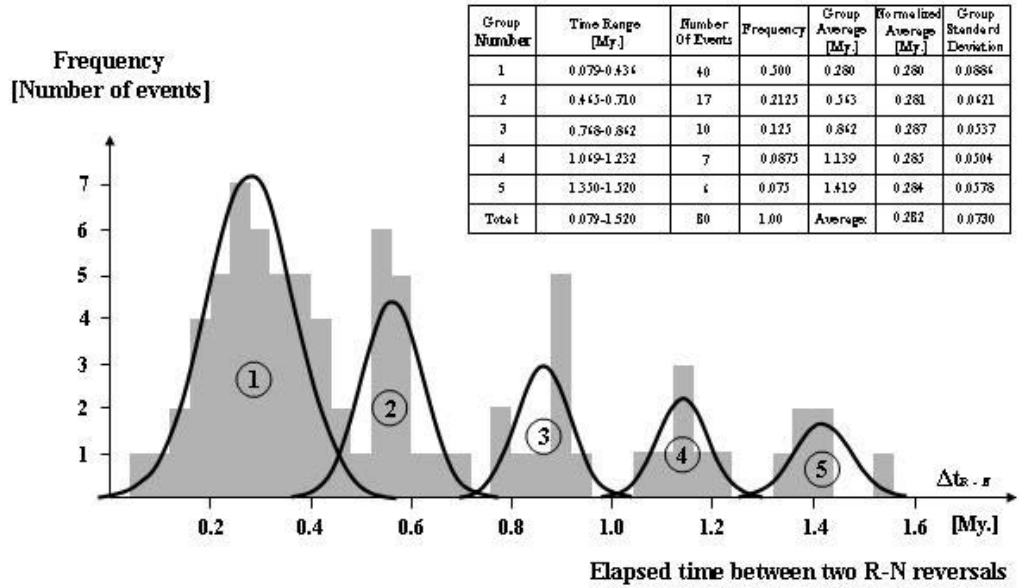

**Figure 1** The distribution of the time intervals elapsed between two consecutive R-N reversals, and the best fitting normal distributions of the separated groups.

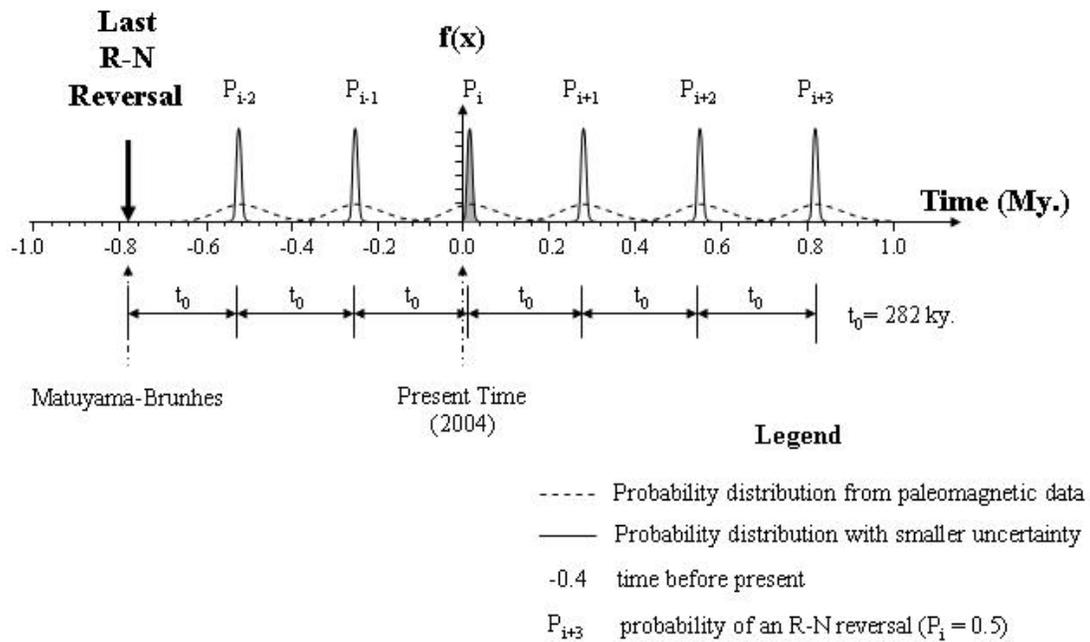

**Figure 2** The probability distribution of the R-N reversals. The calculated statistical parameters contain the relatively high uncertainty of the geologic time determination (dashed lines). If the R-N reversals are generated by astronomical cycles then the uncertainty of the parameters should be significantly smaller (continuous lines).